\documentclass[aps,prd,onecolumn,showpacs,showkeys,amsmath,amssymb]{revtex4}
\usepackage{epsfig} 
\usepackage{amsmath}  
\usepackage{makecell}
\usepackage{graphicx}
\usepackage{array}
\usepackage{subfigure}
\usepackage{bm}
\usepackage{longtable}
\usepackage{booktabs}
\usepackage{appendix}
\usepackage{amsfonts}
\usepackage{amsmath}
\usepackage{dcolumn}
\usepackage{siunitx}
\usepackage{bm}
\usepackage{booktabs}
\usepackage[utf8]{inputenc}
\usepackage{float}
\usepackage{longtable,lscape}
\usepackage{txfonts}
\usepackage{overpic}
\usepackage{amssymb}
\usepackage{indentfirst}
\usepackage{epsfig}
\usepackage{feynmf}   
\usepackage{epstopdf}   
\usepackage{slashed}  
\usepackage{color}
\usepackage{multirow}
\usepackage{xcolor}

\usepackage[colorlinks, citecolor=blue,anchorcolor=red,menucolor=red, linkcolor=red,filecolor=red,runcolor=red,urlcolor=blue,frenchlinks=red]{hyperref}

\makeatletter
\newcommand{\figcaption}{\def\@captype{figure}\caption}
\newcommand{\tabcaption}{\def\@captype{table}\caption}

\newcommand{\Rmnum}[1]{\expandafter\@slowromancap\romannumeral #1@}
\def\hlinewd#1{%
  \noalign{\ifnum0=`}\fi\hrule \@height #1 \futurelet
   \reserved@a\@xhline}
\makeatother

\newcommand\qq{\langle\bar{q}q\rangle}
\newcommand\qss{\langle\bar{s}s\rangle}
\newcommand\GG{\langle GG\rangle}

\newcommand\feynintxy{\int^{x_{\mathrm{max}}}_{x_{\mathrm{min}}}\mathrm{d}x\int^{y_{\mathrm{max}}}_{y_{\mathrm{min}}}\mathrm{d}y}
\newcommand\feynintz{\int^{z_{\mathrm{max}}}_{z_{\mathrm{min}}}\mathrm{d}z}
\newcommand\feynintxyt{\int^{1}_{0}\mathrm{d}x\int^{1}_{0}\mathrm{d}y}
\newcommand\feynintzt{\int^{1}_{0}\mathrm{d}z}
\begin{document}
\title{\vspace{-30mm}Decay and production properties of strange double charm pentaquark\vspace{9mm}}

\author{Zi-Yan Yang$^{1,2,3}$}
\email{yangzh@gcu.edu.cn}
\author{Wei Chen$^{4,5}$}
\email{chenwei29@mail.sysu.edu.cn}

\affiliation{$^1$School of Mechanical Engineering and Robotic Engineering, Guangzhou City University of Technology, Guangzhou 510800, China}
\affiliation{$^2$Key Laboratory of Atomic and Subatomic Structure and Quantum Control (MOE), Guangdong Basic Research Center of Excellence for Structure and Fundamental Interactions of Matter, Institute of Quantum Matter, South China Normal University, Guangzhou 510006, China}
\affiliation{$^3$Guangdong-Hong Kong Joint Laboratory of Quantum Matter, Guangdong Provincial Key Laboratory of Nuclear Science, Southern Nuclear Science Computing Center, South China Normal University, Guangzhou 510006, China}
\affiliation{$^4$School of Physics, Sun Yat-sen University, Guangzhou 510275, China}
\affiliation{$^5$Southern Center for Nuclear-Science Theory (SCNT), Institute of Modern Physics, 
Chinese Academy of Sciences, Huizhou 516000, Guangdong Province, China}

\begin{abstract}
In this work we investigate the decay and production properties of the strange double-charm pentaquark $P_{ccs}^{++}$ with strangeness $S=-1$. Building upon our previous work predicting its $J^P=1/2^-$ molecular configuration, we employ three-point QCD sum rules to calculate its strong decay widths and estimate its production branching ratio via $\Xi_{bc}^+$ baryon decays. The total strong decay width to the $\Xi_{cc}\bar{K}$ and $\Omega_{cc}\pi$ final-state channels is determined as $85\pm19$ MeV. Furthermore, using a rescattering mechanism, we analyze the $\Xi_{bc}^+\rightarrow D_s^{\ast-}\Xi_{cc}^{++}\rightarrow D^-P_{ccs}^{++}$ process and estimate the production branching ratio to be $\mathcal{B}r(\Xi_{bc}^+\rightarrow D^-P_{ccs}^{++})=(4.3_{-1.5}^{+2.0})\times10^{-6}$. The relatively narrow width and detectable branching ratio suggest the possibility searching for this pentaquark state in the future.

\end{abstract}
\maketitle

\pagenumbering{arabic}
\section{Introduction}\label{Sec:Intro}
\par The study of exotic multiquark states, proposed early in 1964~\cite{Gell-Mann:1964ewy,1964-Zweig-p-}, has become a pivotal frontier in hadronic physics, offering profound insights into the nonperturbative dynamics of quantum chromodynamics (QCD)~\cite{Nielsen:2009uh,Chen:2016qju,Richard:2016eis,Esposito:2016noz,Ali:2017jda,Guo:2017jvc,Albuquerque:2018jkn,Liu:2019zoy,Brambilla:2019esw,Richard:2019cmi,Faustov:2021hjs,Chen:2022asf,Meng:2022ozq}. Since the discovery of the first hidden-charm pentaquarks $P_c(4380)$ and $P_c(4450)$ by the LHCb Collaboration~\cite{LHCb:2015yax}, significant theoretical and experimental efforts have been devoted to unraveling the nature of these states, which lie beyond the conventional quark model. The recent observation of the double-charm tetraquark $T_{cc}^+(3875)$~\cite{LHCb:2021vvq,LHCb:2021auc} and the strange-charm tetraquark $T_{c\bar{s}}(2900)$~\cite{LHCb:2022lzp} further highlights the rich spectrum of exotic hadrons and underscores the potential existence of double-heavy counterparts in the baryon sector, such as double-charm pentaquarks.

Theoretical attempts have been made to study the mass spectrum from both the hadronic molecular picture~\cite{Yan:2018zdt,Dong:2021bvy,Chen:2017vai,Guo:2017vcf,Zhu:2019iwm,Wang:2023eng,Duan:2024uuf,Wang:2023aob,Wang:2023ael,Sheng:2024hkf} and the compact pentaquark picture~\cite{Chen:2021kad,Xing:2021yid,Zhou:2018bkn,Wang:2018lhz,Park:2018oib}, as well as their electromagnetic properties~\cite{Ozdem:2022vip, Ozdem:2024yel,Zhou:2022gra,Wang:2023ael,Sheng:2024hkf,Zhu:2025abk}. In our previous work~\cite{Yang:2024okq}, we systematically investigated the mass spectra of strange double-charm pentaquarks with quark content $ccus\bar{d}$ and strangeness $S = -1$, employing QCD sum rules for both molecular and compact configurations. Among the predicted configurations, the $J^P = 1/2^-$ molecular pentaquark with $\Xi_{cc}\bar{K}$ structure stands out: its mass (4.20 GeV) lies slightly above the $\Xi_{cc}\bar{K}$ threshold, allowing strong decays only into $\Xi_{cc}\bar{K}$ and $\Omega_{cc}\pi$ channels. This suggests a relatively narrow resonance that could manifest as a discernible peak in experimental invariant mass spectra. The proximity to the threshold suppresses the phase space for strong decays, potentially enhancing its experimental detectability.

\begin{figure}[htbp]
\centering
\includegraphics[width=8cm]{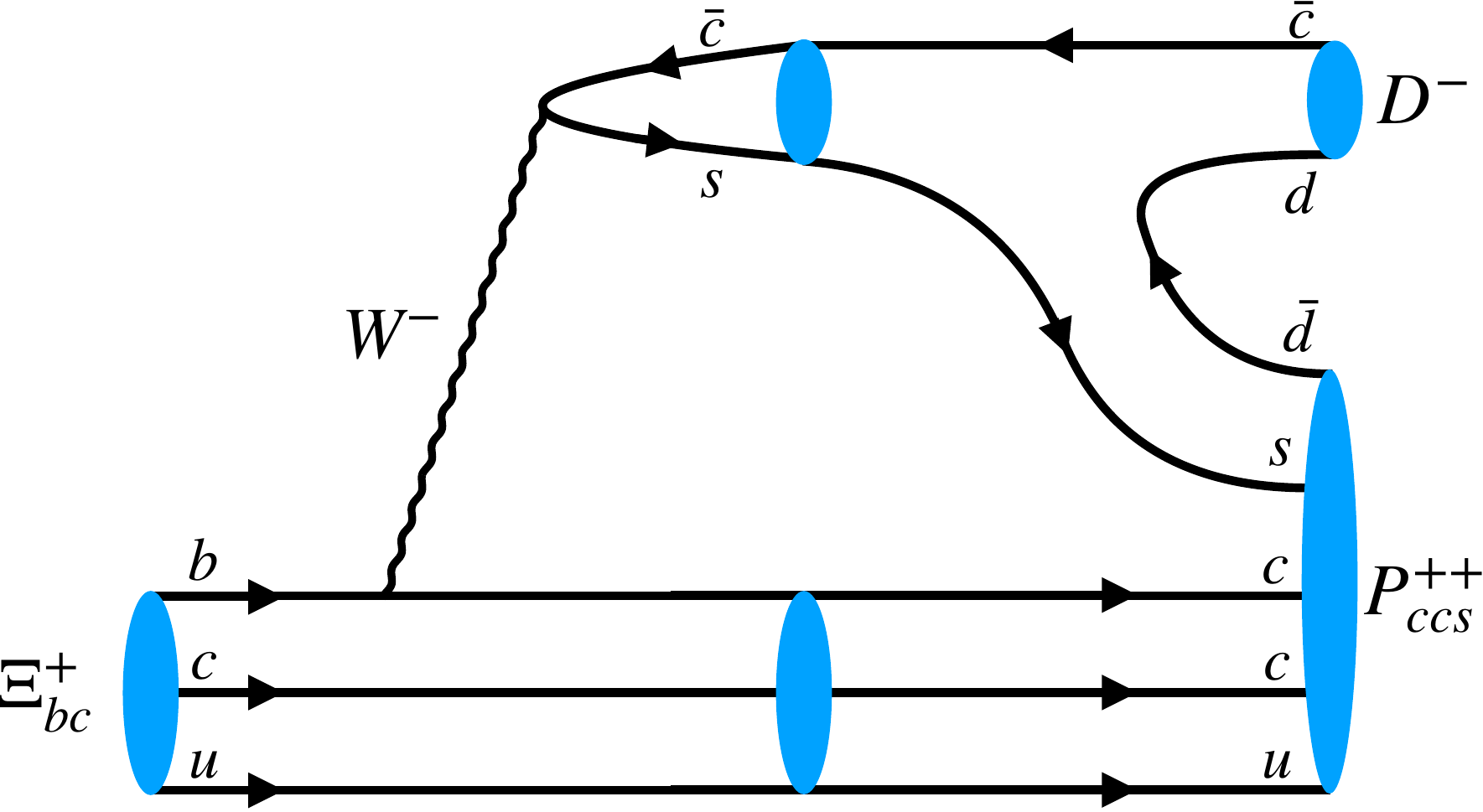}\quad
\includegraphics[width=8cm]{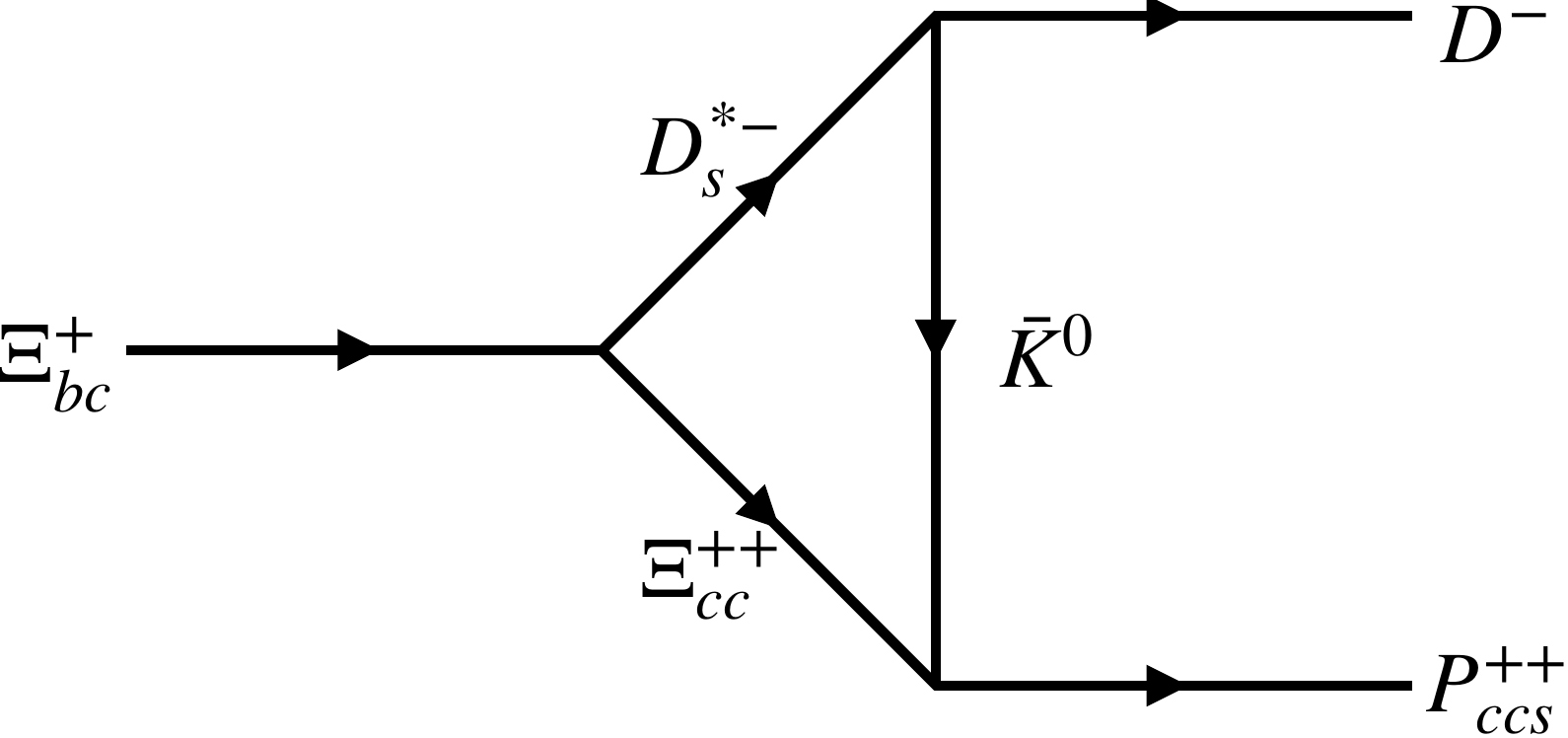}\\
\caption{The production of the strange double-charm pentaquark at the quark level (left) and hadronic level (right).}
\label{Fig:PccsProduction}
\end{figure}

Meanwhile, the ongoing experimental quest for doubly heavy $\Xi_{bc}^+$ baryons offers a pivotal opportunity to unravel the double-charm pentaquark state. Cabibbo-favored decays of $\Xi_{bc}^+$, such as $\Xi_{bc}^+ \to D_s^{*-}\Xi_{cc}^{++}$, could generate the $\Xi_{cc}\bar{K}$ pentaquark through rescattering processes. At the quark level, the diagram for the process $\Xi_{bc}^+\rightarrow D^- P^{++}_{ccs}$ is shown in the left panel of Fig.~\ref{Fig:PccsProduction}. The weak decay arises from a Cabibbo-favored weak transition $b\rightarrow c(\bar{c}s)$ along with the creation of a $d\bar{d}$ pair from the strong interaction. This diagram, known as the external W emission diagram, is non-factorizable because the $s\bar{c}$ pair produced in the weak interaction ends up in different final-state hadrons. Thus, long-distance contributions play a significant role in the $\Xi_{bc}^+\rightarrow D^- P_{ccs}^{++}$ process, where the weakly produced $c\bar{s}$ and $ucc$ pairs hadronize via $\bar{K}^{0}$ meson exchange interaction, the $D_{s}^{\ast-}$ and $\Xi_{cc}^{++}$ pair converts into a $D^{-}$ and the pentaquark $P_{ccs}^{++}$. The corresponding interaction at the hadronic level is shown in the right panel of Fig.~\ref{Fig:PccsProduction}. Such a  mechanism for final-state-interaction (FSI) effects has been successfully applied to $D$ meson decays~\cite{Li:2002pj}, $B$ meson decays~\cite{Cheng:2004ru,Lu:2005mx}, and charm baryon decays~\cite{Han:2021azw,Jia:2024pyb}. Recently, this mechanism has also been applied to the production of tetraquark states in $B$ meson decays~\cite{Chen:2020eyu,Yang:2024coj} and of hidden-charm pentaquark states $P_{c}$ in $\Xi_{b}$ decays~\cite{Hsiao:2024szt}. In this work, we apply this mechanism to the production of the double-charm pentaquark via the $\Xi_{bc}^+\rightarrow D^{\ast-}_s \Xi_{cc}^{++}\rightarrow D^- P^{++}_{ccs}$ process with $\bar{K}^0$ exchange.

A precise calculation of the decay width and production branching ratios of the double-charm pentaquark is thus critical to quantify its production rates in such channels and to guide experimental searches at facilities like LHCb and Belle II. 
In Ref.~\cite{Yang:2024okq}, we also predicted the masses of possible compact pentaquark states with the same quark content. Notably, two compact configurations were found with masses of $4.25 \pm 0.4\;\text{GeV}$, which are close to the mass of the molecular state studied here. It is therefore a natural question how their decay widths and production rates compare with those of the molecular configuration. However, our calculations for these particular compact interpolating currents show that the sum rules analyses for the coupling constants $g_{P_{ccs}\Xi_{cc}\bar{K}}$ and $g_{P_{ccs}\Omega_{cc}\pi}$ did not exhibit a sufficiently stable Borel platform within a reasonable $M_B^2$ window, leading unreliable predictions of these couplings. Since the hadron decay widths and production rates are highly sensitive to the hadronic couplings, it is difficult to provide quantitative estimations for these compact pentaquark states within the framework of three-point QCD sum rules. In this work, we focus on the well-behaved molecular configuration to study the decay and production  properties of strange double charm pentaquark state.
We continue our study by calculating the decay width of the double-charm pentaquark using the QCD sum rule method and estimating its production branching ratios in $\Xi_{bc}$ decays. This paper is organized as follows: In Sec.~\ref{Sec:QCDSR}, we outline the formalism for computing decay widths within the QCD sum rule approach. Sec.~\ref{Sec:Numerical} presents the numerical results for the $J^P=1/2^-$ pentaquark’s decay properties. Sec.~\ref{Sec:FSI} discusses its production mechanism via $\Xi_{bc}^+$ decays and estimates detectable branching ratios. A brief summary is presented in Sec.~\ref{Sec:Summary}.

\section{Three Point QCD Sum Rule}\label{Sec:QCDSR}
\par Over past several decades, the method of QCD sum rule has been proven to be very powerful to study hadron properties~\cite{Reinders:1984sr,Shifman:1978bx,Colangelo:2000dp,Narison:2002woh}. In this section, we shall study the three-point correlation function of several two-body strong decay processes $M\rightarrow X+Y$. For the strong decay process $M\rightarrow X+Y$, the corresponding correlator is written as
\begin{equation}\label{Eq:StrongDecayCF}
\Pi(p,p^{'},q)=\int\mathrm{d}^4x\mathrm{d}^4y\;\mathrm{e}^{\mathrm{i}p^{'}\cdot x}\mathrm{e}^{\mathrm{i}q\cdot y}\langle0|T\{J_{X}(x)J_Y(y)J^\dagger_M(0)\}|0\rangle,
\end{equation}
where $J_{M(X,Y)}$ is the interpolating current for the initial(final) state. In this section, we shall consider the $P_{ccs}\Xi_{cc}\bar{K}$ and $P_{ccs}\Omega_{cc}\pi$ strong decay vertices with $K(\pi)$ off shell. We use the following interpolating current for $P_{ccs}^{++}$ by considering it as a $\Xi_{cc}\bar{K}$ molecule~\cite{Yang:2024okq}:
\begin{equation}\label{Eq:xi1}
\xi_{1}=[\epsilon_{abc}(c_a^T\mathcal{C}\gamma_\mu c_b)\gamma_\mu\gamma_5u_c][\bar{d}_d\gamma_5s_d],
\end{equation}
where $\mathcal{C}$ denotes the charge conjugate operator, subscript $a\cdots d$ denotes the color index and $u,d,s,c$ denotes the up, down, strange, charm quark field, respectively. This current can couple to the $P_{ccs}^{++}$ state with $J^P=1/2^-$ via
\begin{equation}
\langle0|\xi_{1}|P_{ccs}^{1/2^-}\rangle=\lambda^-_{P_{ccs}}u(p),
\end{equation}
in which the value of the coupling constant $\lambda^-_{P_{ccs}}$ are determined from the two-point mass sum rules established in Ref.~\cite{Yang:2024okq} :
\begin{equation}
\lambda^-_{P_{ccs}}=(2.3\pm0.7)\times10^{-3}\;\mathrm{GeV}^6.
\end{equation}
The interpolating currents for $\bar{K}$ and $\pi^+$ mesons can be constructed as 
\begin{equation}
J_{\bar{K}}=\mathrm{i}\,\bar{d}_{a}\gamma_5s_a,\quad J_{\pi^+}=\mathrm{i}\,\bar{d}_{a}\gamma_5u_a,
\end{equation}
which can couple to the meson states via
\begin{equation}
\langle0|J_{\bar{K}}|\bar{K}\rangle=f_{\bar{K}}\frac{m_K^2}{m_s}\equiv\lambda_K,\quad\langle0|J_{\pi^+}|\pi^+\rangle=f_{\pi}\frac{m_\pi^2}{m_u+m_d}\equiv\lambda_\pi.
\end{equation}
The interpolating currents for double charm baryons are taken as~\cite{Zhang:2009iya}
\begin{equation}
\begin{split}
J_{\Xi_{cc}}&=\epsilon_{abc}(c^T_aC\gamma_\mu c_b)\gamma_\mu\gamma_5u_c,\\ 
J_{\Omega_{cc}}&=\epsilon_{abc}(c^T_aC\gamma_\mu c_b)\gamma_\mu\gamma_5s_c,    
\end{split}
\end{equation}
which can couple to the baryon states via
\begin{equation}
\begin{split}
\langle0|J_{\Xi_{cc}}|\Xi_{cc}(p,s)\rangle&=f_{\Xi_{cc}}u(p,s),\\
\langle0|J_{\Omega_{cc}}|\Omega_{cc}(p,s)\rangle&=f_{\Omega_{cc}}u(p,s).
\end{split}
\end{equation}
The coupling constant $g_{P_{ccs}\Xi_{cc}\bar{K}}$ and $g_{P_{ccs}\Omega_{cc}\pi}$ are defined via the effective Lagrangian~\cite{Zou:2002yy}
\begin{equation}
\begin{split}
\mathcal{L}_{P_{ccs}\Xi_{cc}\bar{K}}&=g_{P_{ccs}\Xi_{cc}\bar{K}}P_{ccs}\bar{\Xi}_{cc}\bar{K},\\
\mathcal{L}_{P_{ccs}\Omega_{cc}\pi}&=g_{P_{ccs}\Omega_{cc}\pi}P_{ccs}\bar{\Omega}_{cc}\bar{\pi},
\end{split}
\end{equation}
thus the transition matrix element can be obtained as
\begin{equation}
\begin{split}
\langle\Xi_{cc}(p^{'},s^{'})\bar{K}(q)|P_{ccs}(p)\rangle&=g_{P_{ccs}\Xi_{cc}\bar{K}}\bar{u}_{\Xi_{cc}}(p^{'},s^{'})u_{P_{ccs}}(p,s), \\
\langle\Omega_{cc}(p^{'},s^{'})\pi(q)|P_{ccs}(p)\rangle&=g_{P_{ccs}\Omega_{cc}\pi}\bar{u}_{\Omega_{cc}}(p^{'},s^{'})u_{P_{ccs}}(p,s). 
\end{split}
\end{equation}
With the above coupling relations and transition matrix element, we can obtain the three-point correlation function Eq.~\eqref{Eq:StrongDecayCF} for $P_{ccs}^{++}\rightarrow \Xi_{cc}\bar{K}$ on the phenomenological side 
\begin{equation}
\begin{split}
\Pi(p,p^{'},q)=&\int\mathrm{d}^4x\mathrm{d}^4y\;\mathrm{e}^{\mathrm{i}p^{'}\cdot x}\mathrm{e}^{-\mathrm{i}q\cdot y}\langle0|T\{J_{P_{ccs}}(x)J^\dagger_{K}(y)J^\dagger_{\Xi_{cc}}(0)\}|0\rangle\\
=&\frac{\lambda_{P_{ccs}}^{-}\lambda_{\Xi_{cc}}\lambda_{K}g_{P_{ccs}\Xi_{cc}\bar{K}}}{(p^{2}-m_{P_{ccs}}^2)(p^{'2}-m_{\Xi_{cc}}^2)(q^2-m_K^2)}(\slashed{p}+m_{P_{ccs}})(\slashed{p}^{'}+m_{\Xi_{cc}})+\cdots,
\end{split}
\end{equation}
and for $P_{ccs}^{++}\rightarrow \Omega_{cc}\pi$ 
\begin{equation}
\begin{split}
\Pi(p,p^{'},q)=&\int\mathrm{d}^4x\mathrm{d}^4y\;\mathrm{e}^{\mathrm{i}p^{'}\cdot x}\mathrm{e}^{-\mathrm{i}q\cdot y}\langle0|T\{J_{P_{ccs}}(x)J^\dagger_{\pi}(y)J^\dagger_{\Omega_{cc}}(0)\}|0\rangle\\
=&\frac{\lambda_{P_{ccs}}^{-}\lambda_{\Xi_{cc}}\lambda_{K}g_{P_{ccs}\Omega_{cc}\pi}}{(p^{2}-m_{P_{ccs}}^2)(p^{'2}-m_{\Omega_{cc}}^2)(q^2-m_{\pi}^2)}(\slashed{p}+m_{P_{ccs}})(\slashed{p}^{'}+m_{\Omega_{cc}})+\cdots,
\end{split}
\end{equation}
On the OPE side, we can evaluate the correlation function with standard QCD sum rule approach. To establish a sum rule for the coupling constant, we will pick out the $1/Q^2$ terms around the pole $Q^2\sim0$ (where $Q^2=-q^2$) with the structure $\slashed{p}$ in the OPE series and then match both sides of the sum rule. 
The mass of the off-shell meson ($\bar{K}$ or $\pi$) is much smaller than the masses of the baryons ($\Xi_{cc}$, $\Omega_{cc}$, $P_{ccs}$). This justifies the approximation of extracting the $1/q^2$ pole term directly in the QCD sum rule analysis, making an exponential extrapolation to the on-shell point unnecessary.
After performing the Borel transform $P^2\rightarrow M_B^2$ on both phenomenological and OPE sides, we obtain the strong coupling for the $P_{ccs}\Xi_{cc}\bar{K}$ vertex
\begin{equation}\label{Eq:StrongCoupling1}
g_{P_{ccs}\Xi_{cc}\bar{K}}(s_0,M_B^2)=\frac{1}{\lambda_{P_{ccs}}^{-}\lambda_{\Xi_{cc}}\lambda_{K}(m_{P_{ccs}}+m_{\Xi_{cc}})}\frac{m_{P_{ccs}}^2-m_{\Xi_{cc}}^2}{\mathrm{e}^{-m_{\Xi_{cc}}^2/M_B^2}-\mathrm{e}^{-m_{P_{ccs}}^2/M_B^2}}\left(\int_{s_<}^{s_0}\mathrm{d}s\;\rho(s)\mathrm{e}^{-s/M_B^2}+R(M_B^2)\right),
\end{equation}
and strong coupling for $P_{ccs}\Omega_{cc}\pi$ vertex
\begin{equation}\label{Eq:StrongCoupling2}
g_{P_{ccs}\Omega_{cc}\pi}(s_0,M_B^2)=\frac{1}{\lambda_{P_{ccs}}^{-}\lambda_{\Omega_{cc}}\lambda_{\pi}(m_{P_{ccs}}+m_{\Omega_{cc}})}\frac{m_{P_{ccs}}^2-m_{\Omega_{cc}}^2}{\mathrm{e}^{-m_{\Omega_{cc}}^2/M_B^2}-\mathrm{e}^{-m_{P_{ccs}}^2/M_B^2}}\left(\int_{s_<}^{s_0}\mathrm{d}s\;\rho(s)\mathrm{e}^{-s/M_B^2}+R(M_B^2)\right),
\end{equation}
where the continuum threshold $s_0=22.3\;\mathrm{GeV}^2$ is taken from the two-point mass sum rules in Ref.~\cite{Yang:2024okq}. 
\par Using the operator product expansion (OPE) method, the three-point function can also be evaluated at the quark-gluonic level as a function of various QCD parameters. To evaluate the Wilson coefficients, we adopt the heavy quark propagator in momentum space and the light quark propagator in coordinate space
\begin{eqnarray}
 i S_{Q}^{a b}(p)&=&\frac{i \delta^{a b}}{\slashed{p}-m_{Q}}
 +\frac{i}{4} g_{s} \frac{\lambda_{a b}^{n}}{2} G_{\mu \nu}^{n} \frac{\sigma^{\mu \nu}\left(\slashed{p}+m_{Q}\right)+\left(\slashed{p}+m_{Q}\right) \sigma^{\mu \nu}}{(p^2-m_Q^2)^2}
 +\frac{i \delta^{a b}}{12}\left\langle g_{s}^{2} G G\right\rangle m_{Q} \frac{p^{2}+m_{Q} \slashed{p}}{(p^{2}-m_{Q}^{2})^{4}}, \\
\nonumber i S_{q}^{ab}(x)&=&\frac{i\delta^{ab}}{2\pi^2x^4}\slashed{x}-\frac{\delta^{ab}}{12}\langle\bar{q}q\rangle+\frac{i}{32\pi^2}\frac{\lambda^n_{ab}}{2}g_sG^n_{\mu\nu}\frac{1}{x^2}(\sigma^{\mu\nu}\slashed{x}+\slashed{x}\sigma^{\mu\nu})\\
 & &+\frac{\delta^{ab}x^2}{192}\langle\bar{q}g_s\sigma\cdot Gq\rangle-\frac{m_q\delta^{ab}}{4\pi^2x^2}+\frac{i\delta^{ab}m_q\langle\bar{q}q\rangle}{48}\slashed{x}-\frac{im_q\langle\bar{q}g_s\sigma\cdot Gq\rangle\delta^{ab}x^2\slashed{x}}{1152},
\end{eqnarray}
where $Q$ represents the heavy quark $c$ or $b$, $q$ represents the light quark $u,d,s$, the superscripts $a, b$ denote the color indices. In this work, we will evaluate Wilson coefficients of the correlation function up to dimension nine condensates at the leading order in $\alpha_s$. The spectral function $\rho(s)$ in Eqs.~\eqref{Eq:StrongCoupling1},~\eqref{Eq:StrongCoupling2} are given in Appendix~\ref{Apx:spectrum}. We shall discuss the detail to obtain suitable parameter working regions in QCD sum rule analysis in next section.

\section{Numerical Analysis}\label{Sec:Numerical}
\par In this section we perform the three-point QCD sum rule analysis for double heavy molecular pentaquark systems using the interpolating currents in Eq.~\eqref{Eq:xi1}. We use the standard values of various QCD condensates as $\langle \bar{q}q\rangle(1\mathrm{GeV})=-(0.24\pm0.03)^3\;\mathrm{GeV}^3$, $\langle \bar{q}g_s\sigma\cdot Gq\rangle(1\mathrm{GeV})=-M_0^2\langle \bar{q}q\rangle$, $M_0^2=(0.8\pm0.2)\;\mathrm{GeV}^2$, $\langle \bar{s}s\rangle/\langle \bar{q}q\rangle=0.8\pm0.1$, $\langle g_s^2GG\rangle(1\mathrm{GeV})=(0.48\pm0.14)\;\mathrm{GeV}^4$ at the energy scale $\mu=1$GeV~\cite{Narison:2002woh,Jamin:2001zr,Jamin:1998ra,Ioffe:1981kw,Chung:1984gr,Dosch:1988vv,Khodjamirian:2011ub,Francis:2018jyb} and $m_s(2\;\mathrm{GeV})=95^{+9}_{-3}\;\mathrm{MeV}$, $m_c(m_c)=1.27^{+0.03}_{-0.04}\;\mathrm{GeV}$, $m_b(m_b)=4.18_{-0.03}^{+0.04}\;\mathrm{GeV}$ from the Particle Data Group\cite{ParticleDataGroup:2022pth}. We also take into account the energy-scale dependence of the above parameters from the renormalization group equation~\cite{Wang:2025sic}
\allowdisplaybreaks{
\begin{eqnarray}\label{inputparameter}
\nonumber&&m_s(\mu)=m_s(2\mathrm{GeV})\left[\frac{\alpha_s(\mu)}{\alpha_s(2\mathrm{GeV})}\right]^{\frac{12}{33-2n_f}},\\
\nonumber&&m_c(\mu)=m_c(m_c)\left[\frac{\alpha_s(\mu)}{\alpha_s(m_c)}\right]^{\frac{12}{33-2n_f}},\\
\nonumber&&m_b(m_b)=m_b(m_b)\left[\frac{\alpha_s(\mu)}{\alpha_s(m_b)}\right]^{\frac{12}{33-2n_f}},\\
\nonumber&&\langle \bar{q}q\rangle(\mu)=\langle \bar{q}q\rangle(1\mathrm{GeV})\left[\frac{\alpha_s(1\mathrm{GeV})}{\alpha_s(\mu)}\right]^{\frac{12}{33-2n_f}},\\
  &&\langle \bar{s}s\rangle(\mu)=\langle \bar{s}s\rangle(1\mathrm{GeV})\left[\frac{\alpha_s(1\mathrm{GeV})}{\alpha_s(\mu)}\right]^{\frac{12}{33-2n_f}},\\
\nonumber&&\langle \bar{q}g_s\sigma\cdot Gq\rangle(\mu)=\langle \bar{q}g_s\sigma\cdot Gq\rangle(1\mathrm{GeV})\left[\frac{\alpha_s(1\mathrm{GeV})}{\alpha_s(\mu)}\right]^{\frac{2}{33-2n_f}},\\
\nonumber&&\langle \bar{s}g_s\sigma\cdot Gs\rangle(\mu)=\langle \bar{s}g_s\sigma\cdot Gs\rangle(1\mathrm{GeV})\left[\frac{\alpha_s(1\mathrm{GeV})}{\alpha_s(\mu)}\right]^{\frac{2}{33-2n_f}},\\  
\nonumber&&\alpha_s(\mu)=\frac{1}{b_0t}\left[1-\frac{b_1}{b_0}\frac{\mathrm{log}t}{t}+\frac{b_1^2(\mathrm{log}^2t-\mathrm{log}t-1)+b_0b_2}{b_0^4t^2}\right],
\end{eqnarray}
}
where $t=\mathrm{log}\frac{\mu^2}{\Lambda^2}$, $b_0=\frac{33-2n_f}{12\pi}$, $b_1=\frac{153-19n_f}{24\pi^2}$, $b_2=\frac{2857-\frac{5033}{9}n_f+\frac{325}{27}n_f^2}{128\pi^3}$, $\Lambda=$210 MeV, 292 MeV and 332 MeV for the flavors $n_f=$5, 4 and 3, respectively. In this work, we evolve all the input parameters to the energy scale $\mu=m_c$ for our sum rule analysis. The parameters adopted for the $K$ and $\pi$ mesons and double charm baryons are displayed in Tab.~\ref{Tab:Hadronicinput}.
\begin{table}[htbp]
\caption{The values of the hadronic parameters $m_H$ and $f_H$ in the work taken from Refs.~\cite{ParticleDataGroup:2022pth,Brown:2014ena, Wang:2010hs, Shi:2019hbf}.}\label{Tab:Hadronicinput}\renewcommand\arraystretch{1.6} 
\setlength{\tabcolsep}{0.5 em}{ 
\centering
\begin{tabular}{c c c|c c c}
  \hline
 \hline
Meson(M) & Mass $m_{M}$[GeV] & Decay constant $f_M$[GeV] & Baryon(B) &  Mass $m_B$[GeV] & Decay constant $f_B$[GeV] \\
\hline
$\pi$ & 0.140 & $0.16\pm0.04$ & $\Xi_{cc}^{++}$ & 3.621 & 0.109\\
$K$ & 0.494 & $0.16\pm0.02$ & $\Omega_{cc}^{+}$ & 3.738 & 0.138\\
 \hline
  \hline
\end{tabular}}
\end{table}
\subsection{Strong coupling $g_{P_{ccs}\Xi_{cc}\bar{K}}$}
\begin{figure}[htbp]
\centering
\includegraphics[width=8cm]{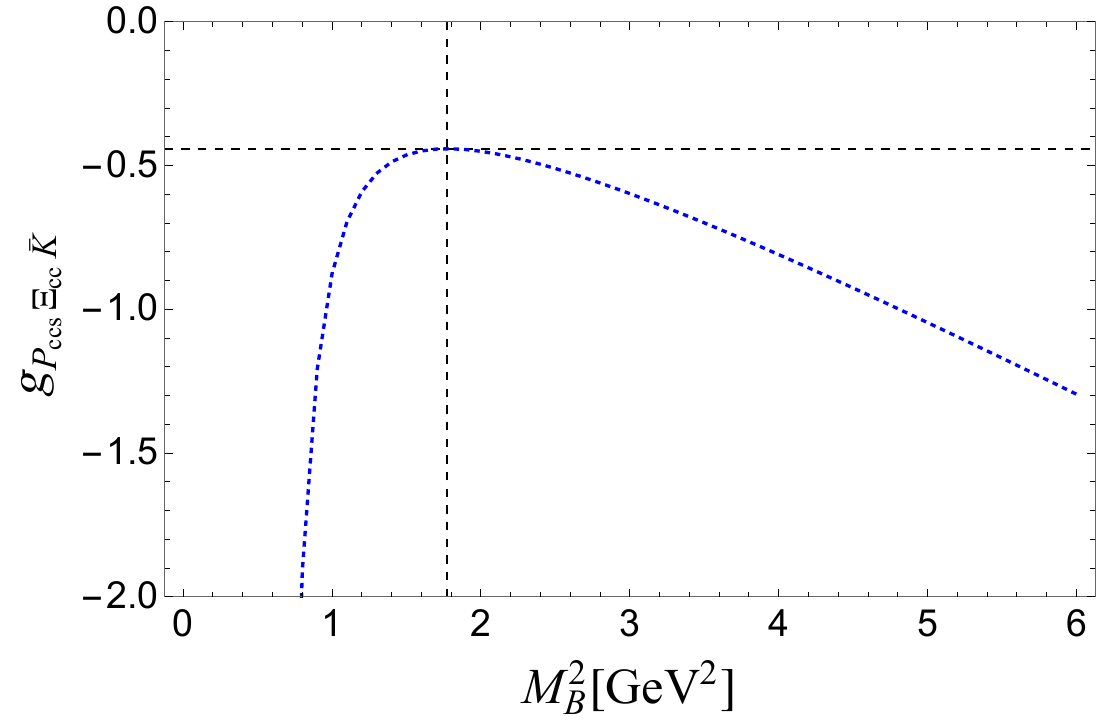}\\
\caption{The dependence of the strong coupling $g_{P_{ccs}\Xi_{cc}\bar{K}}$ on the Borel mass $M_B^2$. The transfer momentum is set to be $Q^2=m_{\Xi_{cc}}^2\sim 13.1\;\mathrm{GeV}^2$.}
\label{Fig:gDKXMB&Q2}
\end{figure}
In Fig.~\ref{Fig:gDKXMB&Q2}, we show the variation of the coupling constant $g_{P_{ccs}\Xi_{cc}\bar{K}}(Q^2)$ with the Borel mass $M_B^2$ at $Q^2=m_{\Xi_{cc}}^2 \approx 13.1\;\text{GeV}^2$. This momentum point is chosen far away from the kaon pole $m_K^2$ so that the approximation of extracting the $1/Q^2$ term in the QCD sum rule is valid. We find that the coupling constant has a maximum value at $ M_B^2 \sim 1.77\;\text{GeV}^2$, where it shows minimal dependence on the unphysical Borel parameter. Since the kaon mass is much smaller than the masses of the baryons $\Xi_{cc}$ and $P_{ccs}$, the $Q^2$-dependence of the coupling constant is negligible in the range from the sum rule working point $Q^2=m_{\Xi_{cc}}^2$ to the physical on-shell point $Q^2 = -m_K^2$. Therefore, we can directly use the coupling constant evaluated at $Q^2=m_{\Xi_{cc}}^2$ to compute the decay width without performing an extrapolation. The extracted coupling constant is:
\begin{equation}\label{Eq:gPXiKresult}
g_{P_{ccs}\Xi_{cc}\bar{K}}(m_{\Xi_{cc}}^2) = -(0.45 \pm 0.05)\;\text{GeV}^{-3}.
\end{equation}
This value is then used to compute the decay width. From the matrix element, we can obtain the decay width for $P_{ccs}^{++}\rightarrow\Xi_{cc}^{++}\bar{K}^0$ process:
\begin{equation}
\Gamma(P_{ccs}^{++}\rightarrow\Xi_{cc}^{++}\bar{K}^0)=\frac{\sqrt{\lambda(m_{P_{ccs}}^2,m_{\Xi_{cc}}^2,m_K^2)}}{8\pi m_{P_{ccs}}^2}g_{P_{ccs}\Xi_{cc}\bar{K}}(m_{\Xi_{cc}}^2)\left((m_{P_{ccs}}+m_{\Xi_{cc}})^2-m_K^2\right).
\end{equation}
Substituting the above coupling, we can obtain the decay width as
\begin{equation}
\Gamma(P_{ccs}^{++}\rightarrow\Xi_{cc}^{++}\bar{K}^0)=65\pm16\,\mathrm{MeV}.
\end{equation}

\subsection{Strong coupling $g_{P_{ccs}\Omega_{cc}\pi}$}
\begin{figure}[htbp]
\centering
\includegraphics[width=8cm]{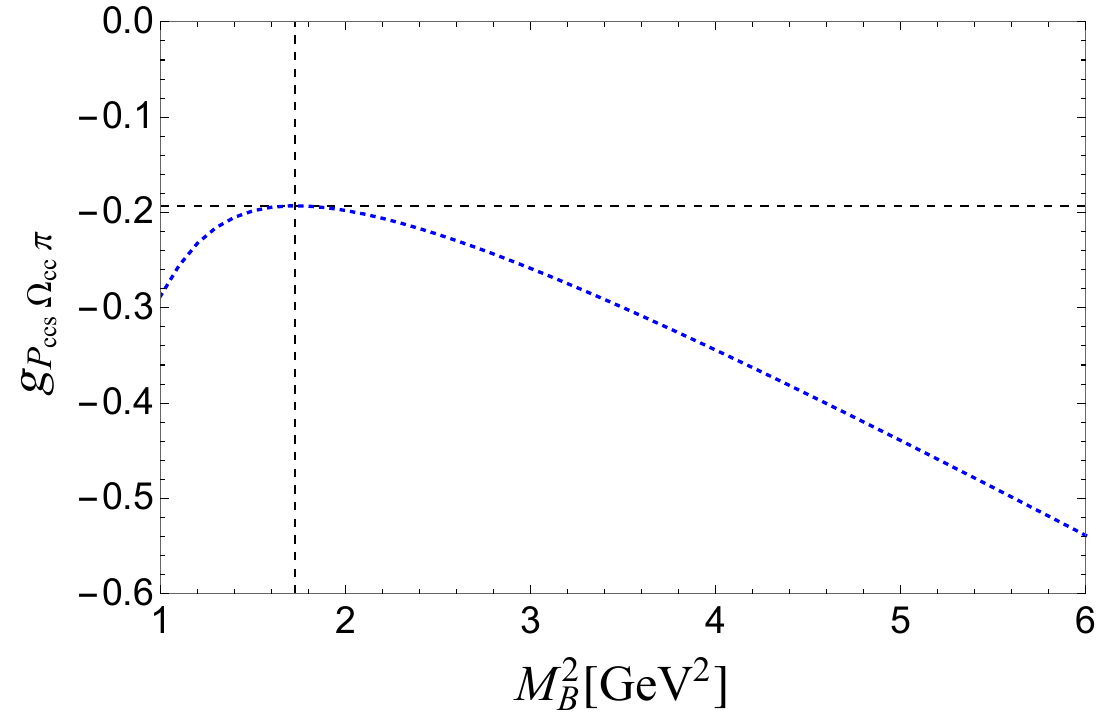}\\
\caption{The dependence of the strong coupling $g_{P_{ccs}\Omega_{cc}\pi}$ on the Borel mass $M_B^2$. The transfer momentum is set to be $Q^2=m_{\Omega_{cc}}^2\sim 13.8\;\mathrm{GeV}^2$.}
\label{Fig:gDsKsXMB&Q2}
\end{figure}
For the $P_{ccs}\Omega_{cc}\pi$ vertex, we show the variation of the coupling constant $g_{P_{ccs}\Omega_{cc}\pi}(Q^{2})$ with the Borel mass $M_{B}^{2}$ at $Q^{2}=m_{\Omega_{cc}}^{2}\approx 13.8\;\text{GeV}^{2}$ in Fig.~3. 
Similar to the discussion above, we extract the $1/Q^{2}$ term in the QCD sum rule since the negligible pion mass. The dependence of the coupling constant on the unphysical Borel parameter achieves minimum around $M_{B}^{2}\sim 1.74\;\text{GeV}^{2}$. Neglecting the small $Q^{2}$-dependence, we compute the coupling constant at $Q^{2}=m_{\Omega_{cc}}^{2}$ as 
\begin{equation}
g_{P_{ccs}\Omega_{cc}\pi}(m_{\Omega_{cc}}^{2}) = -(0.19^{+0.05}_{-0.06})\;\text{GeV}^{-3}.
\end{equation}
From the matrix element, we obtain the decay width for the $P_{ccs}^{++}\to \Omega_{cc}^{+}\pi^{+}$ process:
\begin{equation}
\Gamma(P_{ccs}^{++}\rightarrow\Omega_{cc}^+\pi^+)=\frac{\sqrt{\lambda(m_{P_{ccs}}^2,m_{\Omega_{cc}}^2,m_\pi^2)}}{8\pi m_{P_{ccs}}^2}g_{P_{ccs}\Omega_{cc}\pi}(m_{\Omega_{cc}}^{2})\left((m_{P_{ccs}}+m_{\Omega_{cc}})^2-m_\pi^2\right).
\end{equation}
Substituting the above coupling, we can obtain the decay width as
\begin{equation}
\Gamma(P_{ccs}^{++}\rightarrow\Omega_{cc}^+\pi^+)=20^{+11}_{-10}\,\mathrm{MeV}.
\end{equation}
Thus, we can obtain the total strong decay width as
\begin{equation}
\begin{split}
\Gamma_{P_{ccs}^{++}}&=\Gamma(P_{ccs}^{++}\rightarrow\Xi_{cc}^{++}\bar{K}^0)+\Gamma(P_{ccs}^{++}\rightarrow\Omega_{cc}^+\pi^+)\\
&=(85\pm19)\,\mathrm{MeV}.
\end{split}
\end{equation}
\section{Production via Final-State-Interaction}\label{Sec:FSI}
\par In the framework of rescattering mechanism, the decay $\Xi_{bc}^+\rightarrow P_{ccs}^{++}D^-$ can most likely proceed as $\Xi_{bc}^+\rightarrow D_s^{\ast-} \Xi^{++}_{cc}\rightarrow D^-P_{ccs}^{++}$ with $K^0$ exchange. Under the factorization approach~\cite{Wirbel:1985ji,Bauer:1986bm,Wang:2017mqp}, we can get the decay amplitude of $\Xi_{bc}^+\rightarrow D_s^{\ast} \Xi_{cc}$:
\begin{equation}\label{Eq:Factorization}
\mathcal{A}(\Xi_{bc}^+\rightarrow D_s^{\ast} \Xi_{cc})=\frac{G_F}{\sqrt{2}}V_{cb}V_{cs}a_1\epsilon^{\ast\mu}\bar{u}_{\Xi_{cc}}\left(A_1\gamma_\mu\gamma_5+A_2\frac{p_{\Xi_{cc},\mu}}{m_{\Xi_{bc}}}\gamma_5+B_1\gamma_\mu+B_2\frac{p_{\Xi_{cc},\mu}}{m_{\Xi_{bc}}}\right).
\end{equation}
The above decay amplitudes in the factorization approach are expressed as
\begin{eqnarray}
A_1&=&-\lambda_{D_s^\ast}\left[g_1(m_{D_s^\ast}^2)+g_2(m_{D_s^\ast}^2)\frac{m_{\Xi_{cc}}-m_{\Xi_{bc}}}{m_{\Xi_{bc}}}\right],\label{Eq:A1}\\
A_2&=&-2\lambda_{D_s^\ast}g_2(m_{D_s^\ast}^2),\label{Eq:A2}\\
B_1&=&\lambda_{D_s^\ast}\left[f_1(m_{D_s^\ast}^2)-f_2(m_{D_s^\ast}^2)\frac{m_{\Xi_{cc}}+m_{\Xi_{bc}}}{m_{\Xi_{bc}}}\right],\label{Eq:B1}\\
B_2&=&2\lambda_{D_s^\ast}f_2(m_{D_s^\ast}^2),\label{Eq:B2}
\end{eqnarray}
where $G_F$ is the Fermi constant, $V_{ik}$ is the CKM matrix elements, $a_1$ is the effective Wilson coefficients obtained by the factorization approach~\cite{Buchalla:1995vs}, and $f_{1,2}$ and $g_{1,2}$ are transition form factors of $\Xi_{bc}^+\rightarrow\Xi_{cc}D_s^\ast$ weak decay process. The above form factors can be parametrized as
\begin{equation}
F(Q^2)=\frac{F(0)}{1-\frac{Q^2}{m_\mathrm{fit}^2}+\delta\left(\frac{Q^2}{m_\mathrm{fit}^2}\right)^2},
\end{equation}
where the parameters $F_0$, $m_\mathrm{fit}$ and $\delta$ are taken from Ref.~\cite{Wang:2017mqp} are listed in Tab.~\ref{Tab:FormFactor}.
\begin{table}[h!]
\caption{The values of the parameters $F(0), m_\mathrm{fit}$ and $\delta$ for the form factors in Eqs.~\eqref{Eq:A1}-\eqref{Eq:B2} for $\Xi_{bc}^+\rightarrow D^{\ast}_{s}\Xi_{cc}$ process taken from Ref.~\cite{Wang:2017mqp}.}\label{Tab:FormFactor}\renewcommand\arraystretch{1.6} 
\setlength{\tabcolsep}{0.5 em}{ 
\centering
\begin{tabular}{c c c c | c c c c}
  \hline
 \hline
Form Factor & $F(0)$ & $m_\mathrm{fit}$ & $\delta$ & Form Factor & $F(0)$ & $m_\mathrm{fit}$ & $\delta$ \\
\hline
$f_1$ & 0.550 & 4.45 & 0.43 & $g_1$ & 0.530 & 4.57 & 0.44 \\
$f_2$ & -0.230 & 4.07 & 0.47 & $g_2$ & -0.043 & 3.90 & 0.48 \\
 \hline
  \hline
\end{tabular}}
\end{table}
The amplitude for the $\Xi_{bc}^+\rightarrow D_s^{\ast-} \Xi_{cc}^{++}\rightarrow D^-P_{ccs}^{++}$ process can be written as:
\begin{equation}\label{Eq:Ampa}
\begin{split}
&\mathcal{A}(\Xi_{bc}^{+}\rightarrow D_s^{\ast-} \Xi^{++}_{cc}\rightarrow D^-P_{ccs}^{++})\\
=&\mathrm{i}\frac{G_F}{\sqrt{2}}V_{cb}V_{cs}a_1\int_{-1}^{1} \mathrm{d}\cos\theta \int_{0}^{2\pi} d\phi\frac{|\mathbf{p}_{D_s^\ast}|}{32\pi^2m_{\Xi_{bc}}^2m_{D_s^\ast}^2}\frac{g_{D_s^\ast DK}(-t)g_{P_{ccs}\Xi_{cc}\bar{K}}(-t)}{t-m_K^2}\bar{u}_{P_{ccs}}(p_{P},s_{P})(\slashed{p}_{\Xi_{cc}}+m_{\Xi_{cc}})H\,u_{\Xi_{bc}}(p_{\Xi_{bc}},s_{\Xi_{bc}}),
\end{split}
\end{equation}
where
\begin{equation}
H=-(p_{D}\cdot p_{D_s^\ast})\left(m_{\Xi_{bc}}\slashed{p}_{D_s^\ast}(A_1\gamma_5+B_1)+p_{D_s^\ast}\cdot p_{\Xi_{cc}}(A_2\gamma_5+B_2)\right)+m_{D_s^\ast}^2\left(m_{\Xi_{bc}}\slashed{p}_D(A_1\gamma_5+B_1)+p_{D}\cdot p_{\Xi_{cc}}(A_2\gamma_5+B_2)\right).
\end{equation}
The corresponding decay width can be written as
\begin{equation}\label{Eq:DecayWidth}
\Gamma(\Xi_{bc}^+\rightarrow P_{ccs}^{++}D^-)=\frac{\sqrt{\lambda(m_{\Xi_{bc}}^2,m_D^2,m_{P_{ccs}}^2)}}{16\pi m_{\Xi_{bc}}^3}|\mathcal{A}(\Xi_{bc}^{+}\rightarrow D_s^{\ast-} \Xi^{++}_{cc}\rightarrow D^-P_{ccs}^{++})|^2.
\end{equation}
\par It should be noted that in some works of the final state interaction formalism~\cite{Chen:2020eyu,Hsiao:2024szt}, the decay amplitude contains the form factor $F(t,m)=(\Lambda^2-m_K^2)/(\Lambda^2-t)$ for each strong vertex, which is introduced to compensate the off-shell effect of the exchanged particle at the vertices~\cite{Gortchakov:1995im}. In this work, we can obtain the strong coupling, such $g_{P_{ccs}\Xi_{cc}\bar{K}}$. We take the result of strong coupling $g_{D_s^\ast DK}(Q^2)$ with QCD sum rule formalism as follow~\cite{Yang:2024coj}:
\begin{equation}\label{Eq:gDsDK}
g_{D_s^\ast DK}(Q^2)=(2.8^{+1.3}_{-0.8}\,\mathrm{GeV^{-2}})\mathrm{e}^{-(0.2\pm0.0\,\mathrm{GeV}^{-2})Q^2}.
\end{equation}
With the values of the strong couplings given in Eqs.~\eqref{Eq:gPXiKresult},~\eqref{Eq:gDsDK} and the result of Eq.~\eqref{Eq:DecayWidth}, and the mass of $\Xi_{bc}^+$ taken from the lattice result~\cite{Brown:2014ena}, the decay width of $\Xi_{bc}^+\rightarrow P_{ccs}^{++}D^-$ process can be calculated as 
\begin{equation}
\Gamma(\Xi_{bc}^+\rightarrow P_{ccs}^{++}D^-)=(1.2_{-0.4}^{+0.6})\times10^{-17}\,\mathrm{GeV}.
\end{equation}
Taking the lattice result of lifetime of $\Xi_{bc}^+$~\cite{Brown:2014ena}, the production branching fraction of $\Xi_{bc}^+\rightarrow P_{ccs}^{++}D^-$ process becomes
\begin{equation}
\mathcal{B}r(\Xi_{bc}^+\rightarrow P_{ccs}^{++}D^-)=(4.3_{-1.5}^{+2.0})\times10^{-6}.
\end{equation}

\section{Summary}\label{Sec:Summary}
Based on our previous calculations of mass spectroscopy~\cite{Yang:2024okq}, we further study the decay and production properties of the exotic strange double charm pentaquark state $P_{ccs}^{++}$ with $J^P=1/2^-$. We perform the three-point QCD sum rules to calculate the coupling constants of $P_{ccs}^{++}\rightarrow\Xi_{cc}^{++}\bar{K}^0$ and $P_{ccs}^{++}\rightarrow\Omega_{cc}^+\pi^+$ strong decay processes. The partial decay widths of these two processes are obtained as 
\begin{equation}
\begin{split}
\Gamma(P_{ccs}^{++}\rightarrow\Xi_{cc}^{++}\bar{K}^0)&=65\pm16\,\mathrm{MeV},\\
\Gamma(P_{ccs}^{++}\rightarrow\Omega_{cc}^+\pi^+)&=20^{+11}_{-10}\,\mathrm{MeV},
\end{split}
\end{equation}
yielding a relative branching ratio $\Gamma(P_{ccs}^{++}\rightarrow\Xi_{cc}^{++}\bar{K}^0):\Gamma(P_{ccs}^{++}\rightarrow\Omega_{cc}^+\pi^+)\approx3.3:1$. The total decay width is predicted as 
\begin{equation}
\Gamma_{P_{ccs}^{++}}=85\pm19\,\mathrm{MeV}\, .
\end{equation}

Furthermore, we study the $\Xi_{bc}^+\rightarrow D_s^{\ast-}\Xi_{cc}^{++}\rightarrow D^-P_{ccs}^{++}$ process via the rescattering mechanism to estimate the branching ratio of $\Xi_{bc}^+\rightarrow D^-P_{ccs}^{++}$ as $(4.3_{-1.5}^{+2.0})\times10^{-6}$. This value of branching ratio aligns with the productions of hidden-charm pentaquark states in $\Xi_b$ decays~\cite{Wu:2021caw, Pan:2023hrk, Wu:2024lud}(having the same $b\rightarrow c\bar{c}s$ weak transition), which typically lie around $\mathcal{B}r\sim10^{-6}-10^{-5}$. 
Combining with the decay branching ratio $\mathcal{B}r(P_{ccs}^{++}\rightarrow\Xi_{cc}^{++}\bar{K}^0)$, we obtain $\mathcal{B}r(\Xi_{bc}^+\rightarrow D^-P_{ccs}^{++}\rightarrow D^-\bar{K}^0\Xi_{cc}^{++})\approx\mathcal{B}r(\Xi_{bc}^+\rightarrow D^-P_{ccs}^{++})\mathcal{B}r(P_{ccs}^{++}\rightarrow \Xi_{cc}^{++}\bar{K}^0)=3.3\times10^{-6}$, providing a benchmark for future experiments. 

To date, LHCb has pursued the search of the $\Xi_{bc}^+$ state exploring the invariant mass of $\Lambda_c\pi$, $\Xi_c\pi$ and $DKp$ final states without significant signals~\cite{LHCb:2020iko, LHCb:2021xba, LHCb:2022fbu}. Nevertheless, high $\Xi_{bc}^+$ yields have been expected at future facilities such as MuIC ($\sim 10^{8}$ events/year)~\cite{Zhao:2025kvq}, CEPC/FCC-ee ($\sim10^{7}$ events/year)~\cite{Ma:2025ito}, LHeC ($\sim 10^{5}$ events/year)~\cite{Bi:2017nzv}, and LHCb Run 3 ($\sim 10^4$ events/year)~\cite{Zhang:2022jst}, which will produce a considerable amount of double charm pentaquark states in the future. Our analysis is specifically applied to the molecular interpretation of the $P_{ccs}^{++}$ state. While compact pentaquark configurations with similar masses are also possible~\cite{Yang:2024okq}, a quantitative analysis of their decay properties within the same QCD sum rule framework proved inconclusive due to the instability of the operator product expansion series for the relevant correlation functions. Future studies with alternative theoretical approaches or improved sum rule constructions may help to clarify the properties of such compact states and their potential distinguishability from molecular candidates in experiments.

\section*{ACKNOWLEDGMENTS}
This work is supported by the National Natural Science Foundation of China with Grant Nos.~ 12575153 and 12175318.
\appendix
\section{Spectral function for three-point correlation of strong vertices}\label{Apx:spectrum}
\par The spectral function $\rho(s)$ and $R(M_B^2)$ in Eq.~\eqref{Eq:StrongCoupling1} is shown as follow
\begin{eqnarray}
\nonumber\rho(s)&=&\feynintxy\frac{3}{512 \pi ^5}y\Big(\pi ^2 \GG m_s (2 \qq-\qss) ((x-2) y+1)-\frac{x}{y-1}\left(\GG+16 \pi ^2 m_s (\qss-2 \qq)\right) \\
& &\Delta(x,y,s)\left(3 (x-1) \Delta(x,y,s)+m_c^2+2 s (x-1) (y-1) y\right)\Big),\\
\nonumber R(M_B^2)&=&\feynintxyt\frac{\GG m_s(\qss-2\qq)((x-2) y+1)}{512 \pi ^3 (x-1)^2 (y-1)^2}\Big(s_1 (x-1)^2 (y-1)^2 y-m_c^2 (x (y (4 ((x-1) x+1) y-2 x-5)+3)\\
& &+y-1)\Big)\mathrm{e}^{-s_1/M_B^2},
\end{eqnarray}
where $x_{\mathrm{min}}=0$, $x_{\mathrm{max}}=\frac{1-2\sqrt{m_c^2/s}}{(1-\sqrt{m_c^2/s})^2}$, $y_{\mathrm{min}}=\frac{s(1-x)+m_c^2x-\sqrt{(s(1-x)+m_c^2x)^2-4m_c^2s(1-x)}}{2s(1-x)}$, $y_{\mathrm{max}}=\frac{s(1-x)+m_c^2x+\sqrt{(s(1-x)+m_c^2x)^2-4m_c^2s(1-x)}}{2s(1-x)}$, $\Delta(x,y,s)=-s(1-y)y+\frac{m_c^2(1-xy)}{1-x}$, $s_1=\frac{m_c^2(1-xy)}{(1-x)(1-y)y}$.

The spectral function $\rho(s)$ and $R(M_B^2)$ in Eq.~\eqref{Eq:StrongCoupling2} is shown as follow
\begin{eqnarray}
\nonumber\rho(s)&=&\feynintxy\frac{3}{512 \pi ^5 (y-1)} \GG x y \Delta(x,y,s) \left(-3 (x-1) \Delta(x,y,s)-m_c^2-2 s (x-1) (y-1) y\right)\\
& &+\feynintz\frac{3 \GG m_s \qss (z-1)}{128 \pi ^3}\\
R(M_B^2)&=&\feynintzt\frac{\GG m_s \qss }{256 \pi ^3z}\left(m_c^2 z-s_2 \left(4 z^2-6 z+2\right)\right)\mathrm{e}^{-s_2/M_B^2}
\end{eqnarray}
where $z_{\mathrm{min}}=\frac{1}{2}\left(1-\sqrt{1-4m_c^2/s}\right)$, $z_{\mathrm{max}}=\frac{1}{2}\left(1+\sqrt{1-4m_c^2/s}\right)$, and $s_2=\frac{m_c^2}{(1-z)z}$.

\end{document}